\documentclass{PoS}

\usepackage[english]{babel}
\usepackage{amsmath}
\usepackage{ol,jan,cite}
\usepackage{graphicx}
\usepackage{axodraw}
\def\msbar1{{\overline{\text{MS}\kern-0.05em}\kern0.05em}}

\title{
\vspace*{-1cm}
\begin{minipage}{\textwidth}
\begin{flushright}
\texttt{\footnotesize
Edinburgh 2008/37 \\
PoS(LAT2008)269\\
}
\end{flushright}
\end{minipage}\\[15pt]
Neutral Kaon Mixing Beyond the Standard Model from 2+1 Flavour Domain Wall QCD
}

\ShortTitle{Neutral Kaon Mixing Beyond the SM}

\author{\speaker{Jan Wennekers}\;
         for the RBC and UKQCD collaborations\\
	 School of Physics and Astronomy, \\
	 The University of Edinburgh,\\
	 Edinburgh EH9 3JZ, UK,\\
         E-mail: \email{jwenneke@ph.ed.ac.uk}}

       \abstract{We present preliminary results of a study of $\Delta S = 2$
         matrix elements originating from physics beyond the Standard Model.
         Using $2+1$ flavour Domain Wall Fermions we obtain the
         non-perturbative renormalisation (mixing) matrix in the RI scheme. We
         also discuss plans for the chiral extrapolation of the renormalised
         matrix elements in a partially quenched set-up.}

\FullConference{The XXVI International Symposium on Lattice Field Theory \\
		 July 14 - 19, 2008\\
		 Williamsburg, Virginia, USA}

\begin{document}

\section{Motivation}
\begin{figure}[h]
  \centering
    \parbox{.45\textwidth}{
      \includegraphics[width=.4\textwidth]{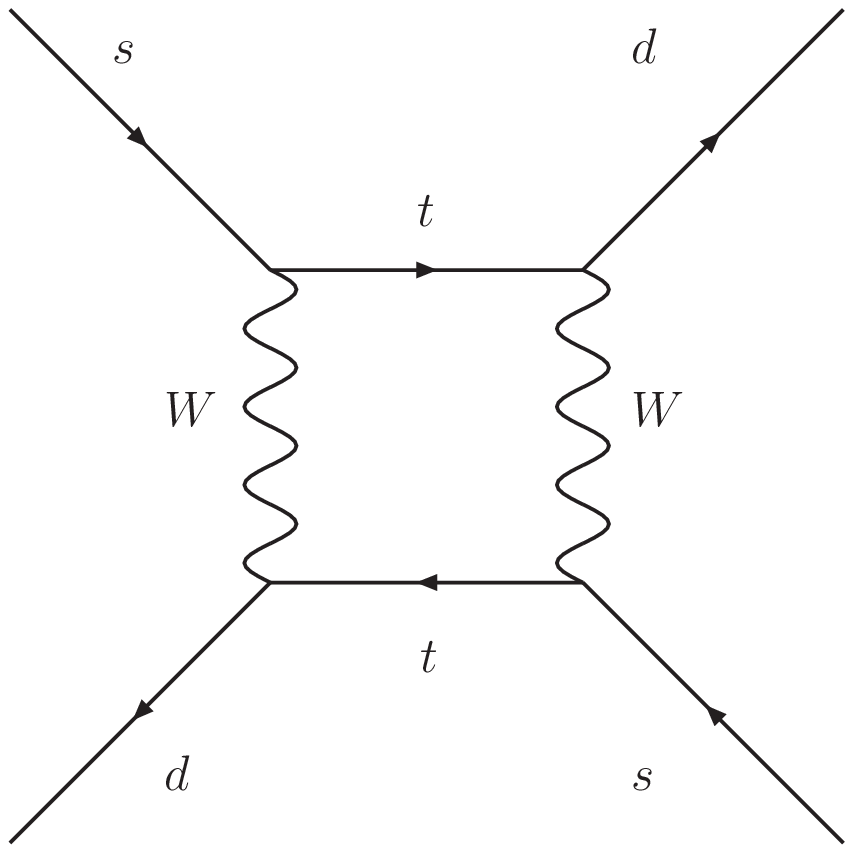}
      \caption{Box diagram for $K_0$-$\ol{K}_0$ mixing in the
	Standard Model.}
      \label{box_sm}}
    \qquad
    \parbox{.45\textwidth}{
      \includegraphics[width=.4\textwidth]{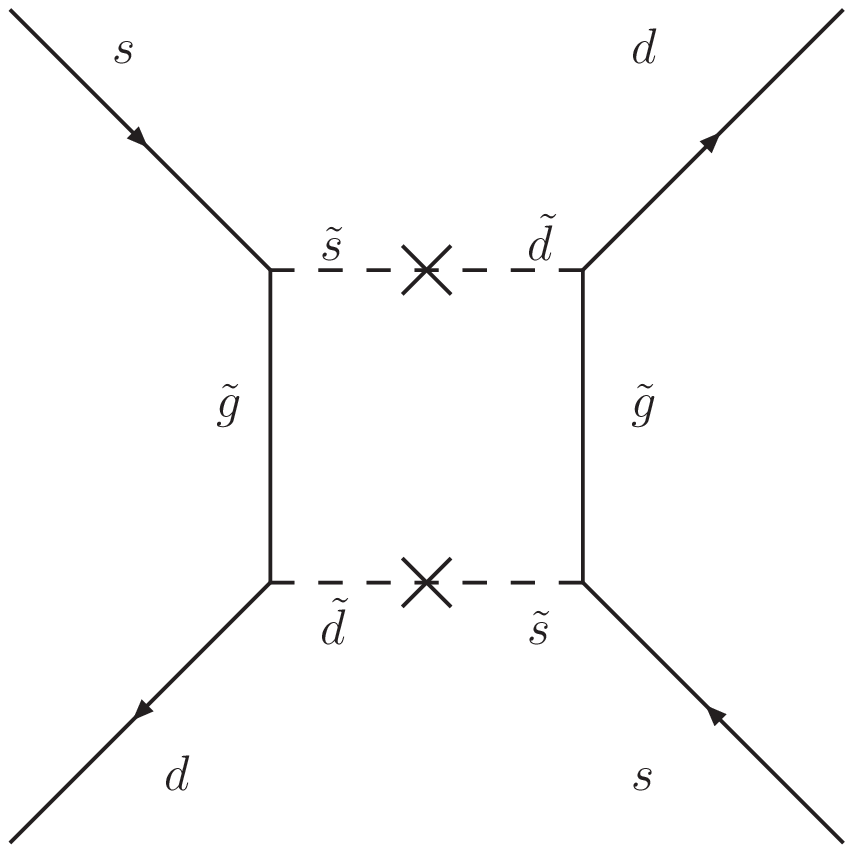}
      \caption{Example for a SUSY box diagram with gluinos and
	squarks in the mass insertion approximation.}
      \label{box_susy}}
\end{figure}

In the Standard Model (SM) there is only one four-quark operator contributing
to the analysis of $K_0-\bar{K}_0$ mixing. It originates from box diagrams
with $W$ bosons, Figure \ref{box_sm}. In theories beyond the SM there are
additional box diagrams. Figure \ref{box_susy} shows an example from the
\emph{mass insertion approximation} of a supersymmetric extension of the SM. 

The matrix element of the SM operator
\begin{align}
   Q_1 &= \ol{s}^a\;\gamma_\mu P_L\;d^a\ol{s}^b\;\gamma^\mu P_L\;d^b,\label{Q1} 
\end{align}
is parametrised by the \emph{bag parameter} $B_K$. $Q_1$ renormalises
multiplicatively for lattice discretisations with chiral symmetry. The most
advanced computations of $B_K$ have exploited the chiral symmetry of Domain
Wall Fermions (DWF) \cite{Antonio:2007pb} or overlap fermions
\cite{Aoki:2008ss}.  The good agreement between the measured value for
$|\epsilon|$ and the SM with lattice $B_K$ input puts a constraint on theories
beyond the Standard Model (BSM) which allow for additional classes of box
diagrams. The lattice can contribute here as well by providing hadronic matrix
elements for a basis of four-quark operators which then can be used in
phenomenological studies. Again lattice fermions with good chiral symmetry are
essential since renormalisation becomes continuum-like.

Besides the SM operator $Q_1$ (\ref{Q1}) we consider the operators, 
\begin{align}
  Q_2 &= \ol{s}^a\;\gamma_\mu P_L\;d^a\ol{s}^b\;\gamma^\mu P_R\;d^b,\\
  Q_3 &= \ol{s}^a\;P_L\;d^a\ol{s}^b\;P_R\;d^b, \\
  Q_4 &= \ol{s}^a\;P_L d^a\ol{s}^b\;P_L\;d^b, \\
  Q_5 &=\ol{s}^a\;\sigma_{\mu\nu}P_L\;d^a\ol{s}^b\;\sigma^{\mu\nu}P_L\;d^b.
\end{align}
This basis is related to the \emph{supersymmetric basis} used in several other
studies by a Fierz transformation. The anomalous dimensions for these
operators are known to next-to-leading order
\cite{Ciuchini:1998ix,Buras:2000if}. The mixing among the BSM operators is
limited to two $2\times2$ blocks,
\begin{align}
  \gamma=
\begin{pmatrix}
  \gamma_{11} & 0 & 0 & 0 & 0 \\
  0 & \gamma_{22} & \gamma_{23} & 0 & 0 \\
  0 & \gamma_{32} & \gamma_{33} & 0 & 0 \\
  0 & 0 & 0 & \gamma_{44} & \gamma_{45} \\
  0 & 0 & 0 & \gamma_{54} & \gamma_{55}
\end{pmatrix}.
\end{align} 

\section{Non-perturbative Renormalisation}
To subtract the logarithmic divergence in the matrix elements of the
operators $Q_i$ we employ the RI scheme \cite{Martinelli:1994ty}
at a scale $\mu$. We consider the full $5\times5$ matrix,
\begin{align}
  Q_i^\text{RI} = Z_{ij}(\mu)Q_j, 
\end{align}
despite the reduced mixing of the continuum theory. Since our lattice
action is chirally symmetric up to a small violation of
$O(m_\text{res})$ we expect the elements of $Z$ which vanish in the
presence of chiral symmetry to be small. Our approach enables us to
check this assumption and to compare the renormalised results with the
full matrix and the block diagonal one.

We compute the amputated four-point vertex functions $\Gamma$ and
project them onto the relevant Dirac structure. The resulting matrix
$\Lambda$ yields the renormalisation matrix, 
\begin{align}
  \frac{1}{Z_q^2} Z(\mu) =\Lambda_\text{tree}\cdot\Lambda^{-1}(p^2=\mu^2).
\end{align}
The wave function renormalisation $Z_q$ is eliminated using the
(local) axial current and its renormalisation constant $Z_A$ which can be
determined independently from the axial Ward identity for Domain Wall
Fermions \cite{Aoki:2007xm}, 
\begin{align}
  \frac{1}{Z_A^2} Z =\Lambda_\text{tree}\cdot\Lambda^{-1}/\Lambda_A.
\end{align}

The standard RI renormalisation condition is defined at an unfortunate
kinematic point. The momenta of all four involved quarks are the same
and there is no momentum flowing out of the vertex. This so called
\emph{exceptional} momentum configuration leads to additional
spontaneous chiral symmetry breaking from subgraphs which are only
suppressed as a small inverse power of momentum. This has been seen in
the splitting between $\Lambda_A$ and $\Lambda_V$, the amputated
vertex functions for the axial and vector current
\cite{Aoki:2007xm}. The effect on the renormalisation factors for the
operators $Q_2$ to $Q_5$ is a $\frac{1}{mp^2}$ divergence.  There are
at least two suggestions in the literature how to deal with this
\emph{pion pole}. One can try to directly fit the pole in
$\Lambda_{ij}$ \cite{Babich:2006bh} or one can form ratios of the type
\cite{Giusti:2000jr}
\begin{align}
\frac{m_1\Lambda_{ij}(m_1,p^2)-m_2\Lambda_{ij}(m_2,p^2)}{m_1-m_2},
\end{align}
where $m_1$ and $m_2$ are two different quark masses. The ratio in the
chiral limit corresponds to the subtracted $\Lambda_{ij}$
\cite{Becirevic:2004ny}.

We have implemented a slightly different renormalisation condition
which directly gives the amputated Green's function without the pion pole. In
this \emph{non-exceptional} case one still has $p_i^2=\mu^2$ for all
legs of the vertex, but there are two pairs of momenta which differ
such that the sum of momenta at the vertex is also $\mu^2$. Now
chirality breaking subgraphs are suppressed by the relatively high
scale.

\begin{figure}[h]
  \centering
  \parbox[t][5cm]{.45\textwidth}{
    \includegraphics[width=.4\textwidth]{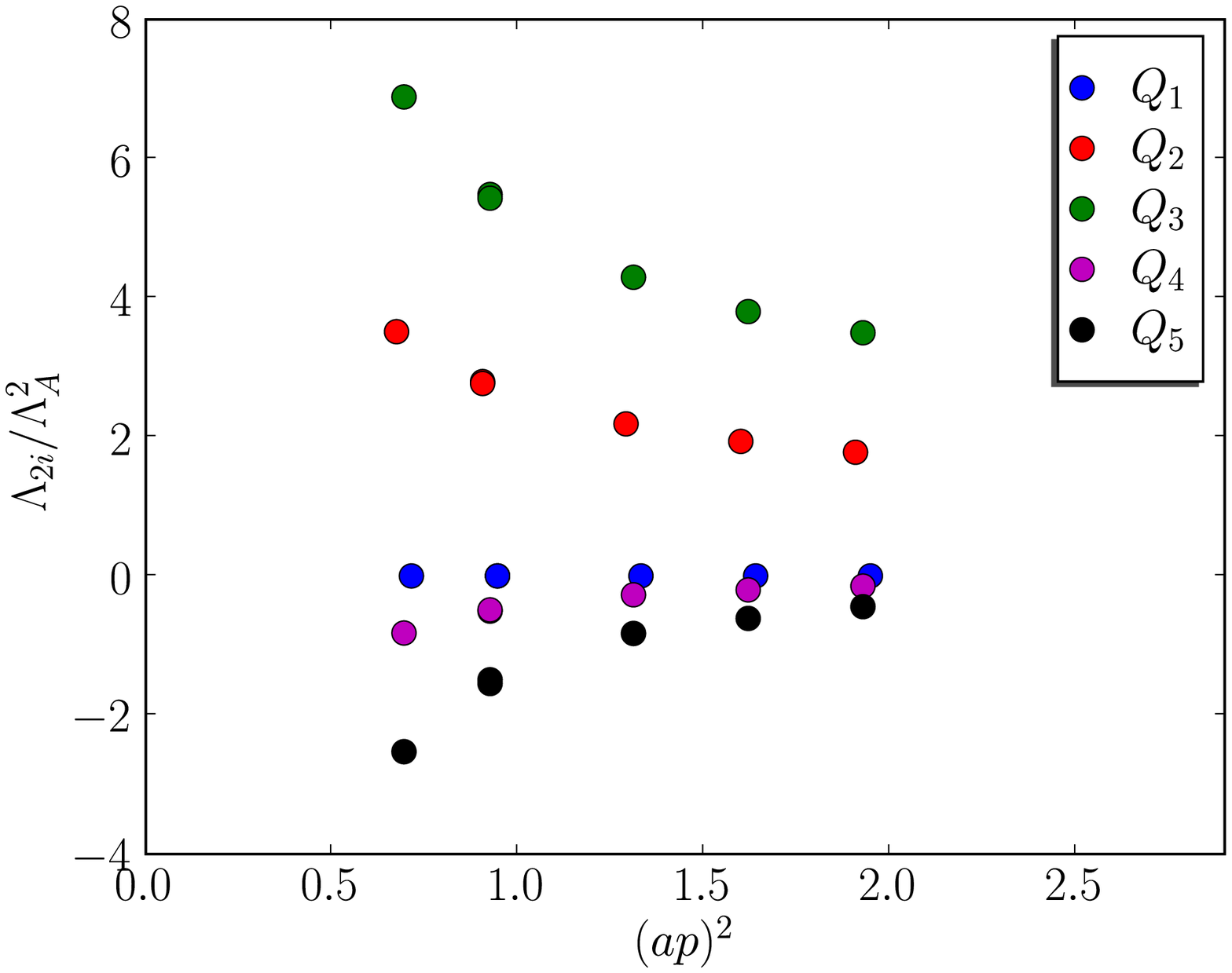}}
  \qquad
  \parbox[t][5cm]{.45\textwidth}{
    \includegraphics[width=.4\textwidth]{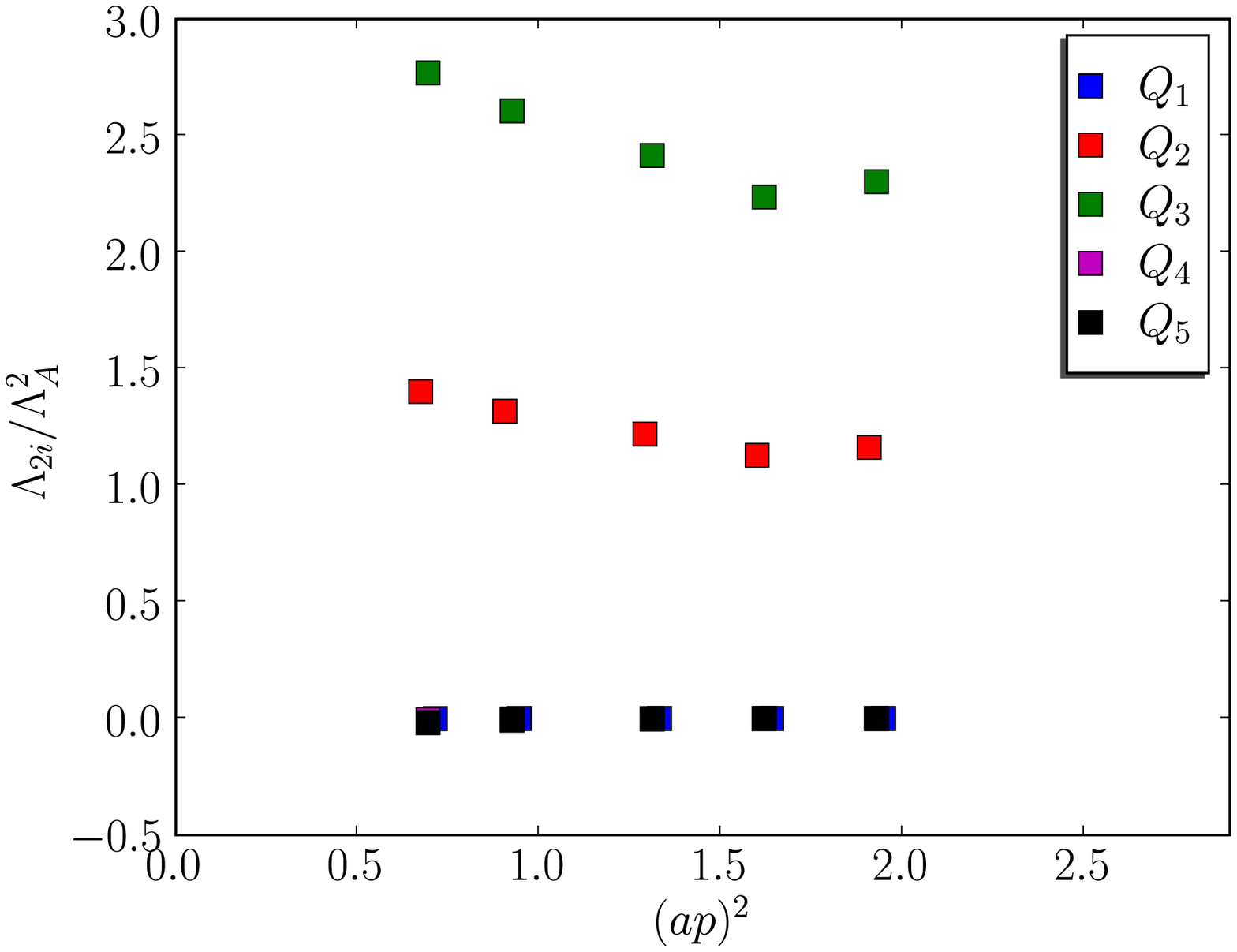}}
  \caption{Elements of the matrix $\Lambda/\Lambda_A^2$ associated
    with the mixing for operator $Q_2$ for a range of $p^2$ with
    exceptional momenta (left) and non-exceptional momenta
    (right). The statistical errors are smaller than the symbols. The
    chiral symmetry of DWF relating $Q_1$ and $Q_2$, $Q_3$ and $Q_4$
    is manifest in the non-exceptional case, but obscured by
    spontaneous chiral symmetry breaking at low $p^2$ in the
    conventional approach.}
  \label{fig_Lambda}
\end{figure}

In Figure \ref{fig_Lambda} we compare the two kinematic set-ups for
the row of the matrix $\Lambda$ which determines the mixing for
$Q_2$. The main difference is the suppression of mixing with the
operators $Q_4$ and $Q_5$ which is caused by the mentioned chirality
breaking subgraphs. This clearly shows the advantages of the
non-exceptional momentum data. The main obstacle for the use of this
new approach is the lack of perturbative results for the matching to
perturbative schemes like $\msbar1$.  A first result has been
presented at this conference for the case of $Z_m$
\cite{Aoki:2008xx}. Such infrared problems not only appear in the
non-perturbtative data, but also seem to influence the rate of
convergence of continuum perturbation theory, which is very promising
for the RI-MOM approach.
 
\vspace{1cm}

Recently we have adopted the use of gauge-fixed momentum sources which project
on a single momentum at the source. The vertex then has to be put at the sink
and a full volume average is possible. These sources allow us to reach a
better statistical accuracy at much reduced cost especially on large volumes.
Even though each momentum requires new propagator inversions the new method
scales much better with the lattice volume and has to be the method of choice
for our new large lattices.

\section{Results for Bare Matrix Elements}

The preliminary results presented in this section are part of the
ongoing project with $32^3\times64\times32$ ensembles of $N_f=2+1$
Domain Wall Fermions with the Iwasaki gauge action at $\beta=2.25$ \cite{Scholz:2008uv}.

\begin{table}[h]
  \begin{center}
    \begin{tabular}{c|c|c|c|c}
      $m_l$ & $m_s$ & $m_\pi$ & Renorm. & Matrix Elements \\ \hline
      & & & & \\
      $0.004$ & $0.03$ & $\sim300\,\text{MeV}$ & $0.004$ & 
      $0.002,0.004,0.006,0.008$ \\
      & & & & $0.025,0.03$ \\
      $0.006$ & $0.03$ & $\sim365\,\text{MeV}$ & $0.006$ & '' \\
      & & & & \\
      $0.008$ & $0.03$ & $\sim420\,\text{MeV}$ & $0.008$ & ''
    \end{tabular}
  \end{center}
\end{table}

The SM matrix element is commonly normalised by its value in the vacuum
saturation approximation. Using the same normalisation for the other four
operators leads to a divergent chiral limit for the resulting bag parameters.
This divergence can be moved into the normalisation, but this requires the
knowledge of renormalised quark masses for the phenomenological use of the
results. Therefore alternative normalisations like are preferable. But for the
purpose of this write-up we stick with the $B_K$ normalisation since all
results are preliminary,
 \begin{align}
    \vev{\bar{K}|Q_i|K}&=N_im_K^2F_K^2B_i,\quad i=1,\ldots,5, \\
    N_i&= \frac{8}{3}, -\frac{4}{3}R, 2R, \frac{5}{3}R, -4R,\label{Nidef} \\
    R &= \left(\frac{m_K}{m_s^r+m_d^r}\right)^2.\label{Rdef}
 \end{align}

\begin{figure}[h]
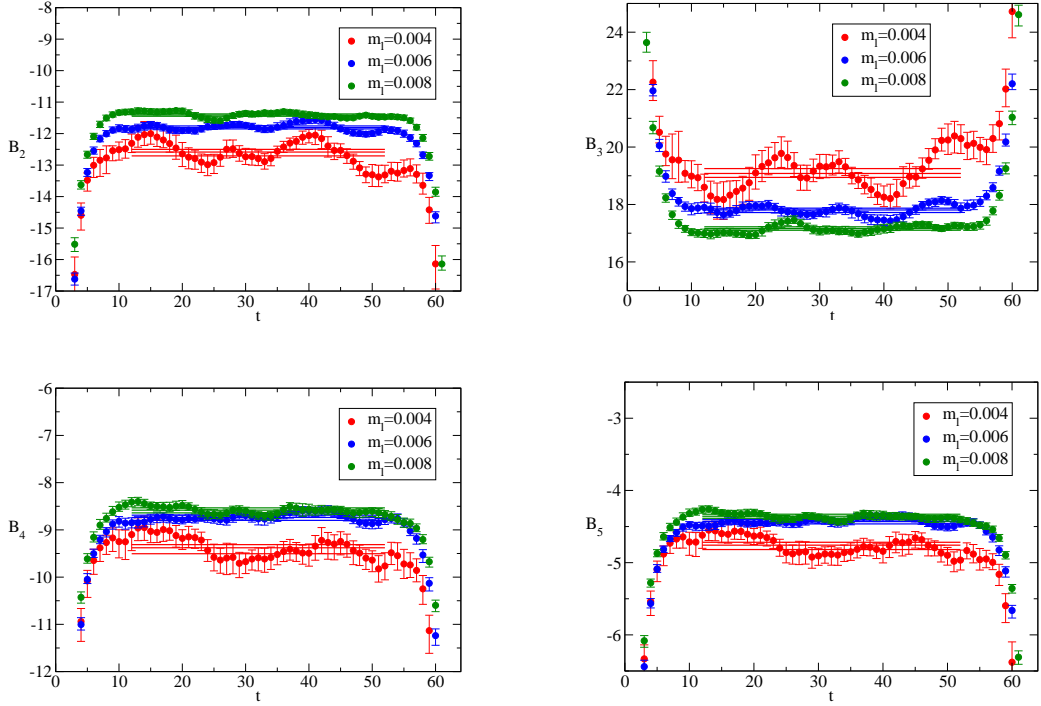

  \centering
  \parbox[t][5cm]{.45\textwidth}{
    \includegraphics[width=.4\textwidth]{B2_0.03_0.004.eps}}
  \qquad
  \parbox[t][5cm]{.45\textwidth}{
    \includegraphics[width=.4\textwidth]{B3_0.03_0.004.eps}}
  \parbox[t][5cm]{.45\textwidth}{
    \includegraphics[width=.4\textwidth]{B4_0.03_0.004.eps}}
  \qquad
  \parbox[t][5cm]{.45\textwidth}{
    \includegraphics[width=.4\textwidth]{B5_0.03_0.004.eps}}
  \caption{Plateau plots for the ratios $B_2$-$B_5$ at the unitary quark masses
  of all three ensembles without normalisation factor $N_i$ (\ref{Nidef}).}
  \label{Bplateaux}
\end{figure}

Gauge-fixed wall sources have been shown to be a very efficient for
$B_K$ \cite{Antonio:2007pb,Kelly:2008aa}. We reach the same level of
statistical uncertainty when applying this approach to the whole
operator basis. In Figure \ref{Bplateaux} we give an impression of the
quality of the data for the four non-SM operators. Each plot shows the
fully dynamical data ($m_\text{val}=m_\text{sea}$) for both light and strange quark
mass. With around 100 configurations the statistical errors are of the
order of 1\% at the lowest dynamical mass for the bare $B$ parameters.
\vspace{1cm}

\begin{figure}[h]
  \centering
  \parbox[t][5cm]{.45\textwidth}{
    \includegraphics[width=.4\textwidth]{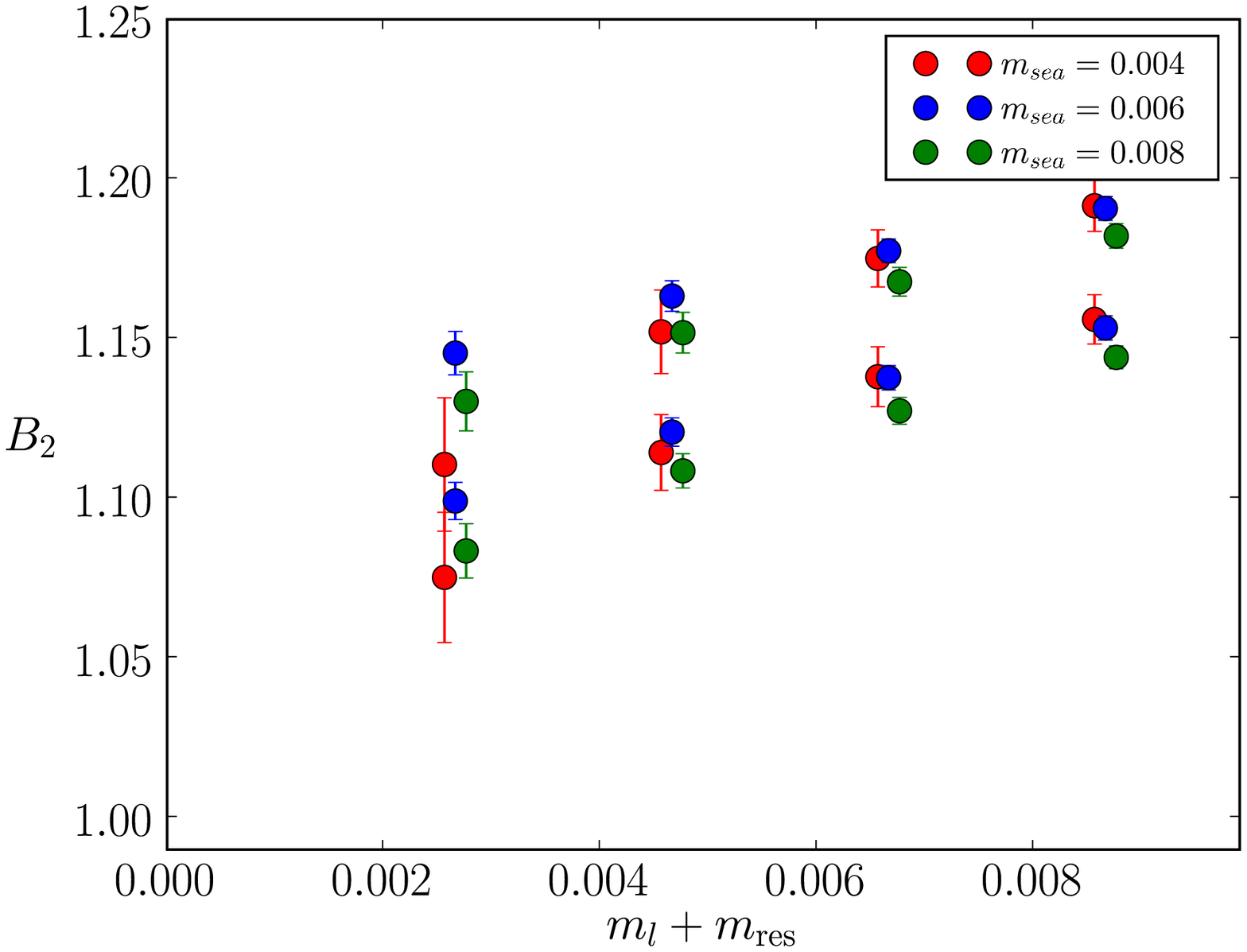}}
  \qquad
  \parbox[t][5cm]{.45\textwidth}{
    \includegraphics[width=.4\textwidth]{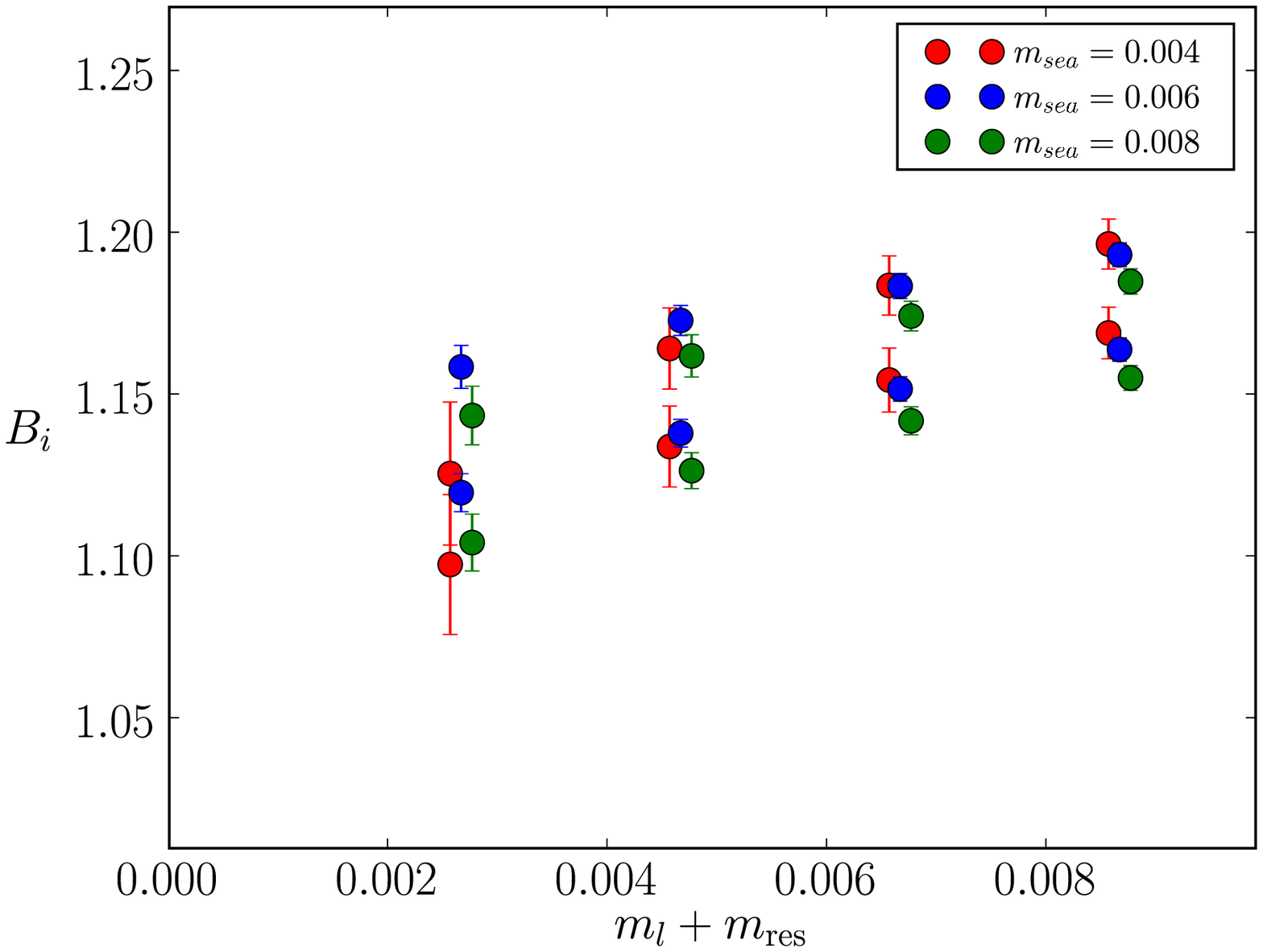}}
  \parbox[t][5cm]{.45\textwidth}{
    \includegraphics[width=.4\textwidth]{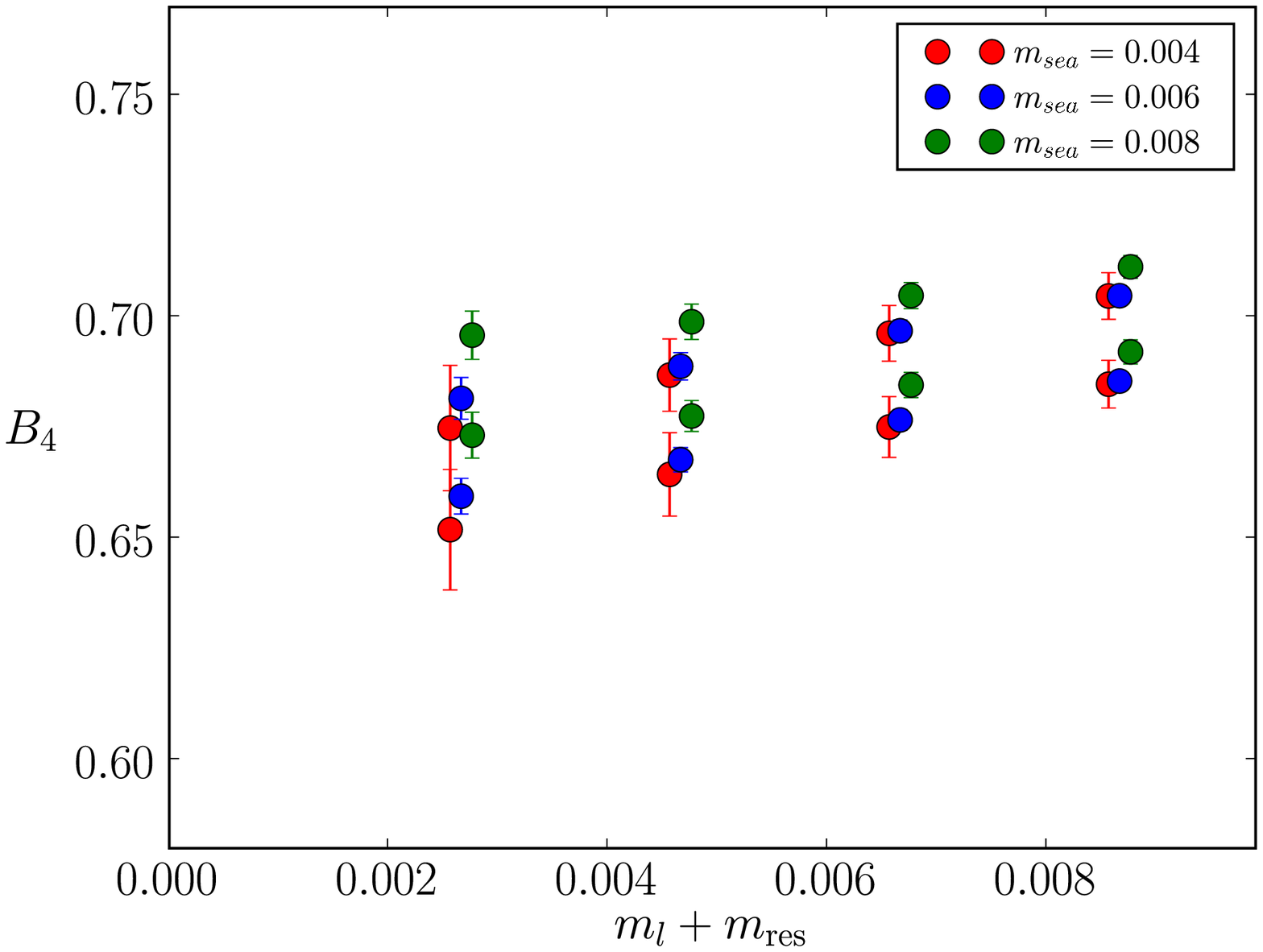}}
  \qquad
  \parbox[t][5cm]{.45\textwidth}{
    \includegraphics[width=.4\textwidth]{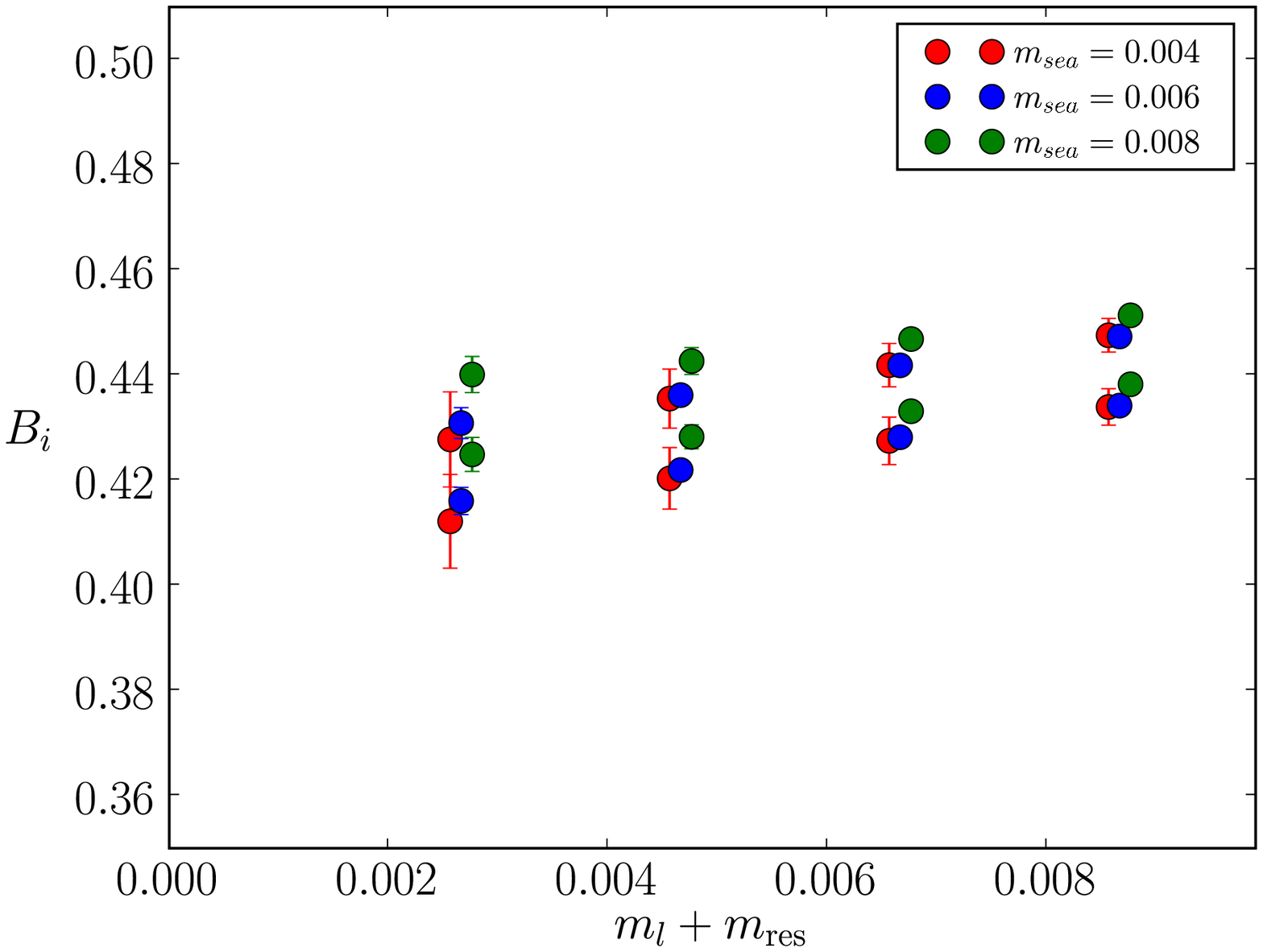}}
  \caption{Light quark mass dependence for the bag parameters
  $B_2$-$B_5$. The different colours indicate the three sea quark
  masses. In each plot there is data for two strange quark masses,
  0.025 and 0.03.}
  \label{fig_chiral}
\end{figure}

Our plan for the chiral extrapolation of this data set is again along
the lines of the existing work on $B_K$. We use an approach based of
$SU(2)$ Chiral Perturbation Theory (ChPT). In this setup the leading
order Low Energy Constants $F$ and $B_0$ are fixed from fits in the
pion sector. The kaon is treated as a heavy meson
\cite{Allton:2008pn}.  The needed partially quenched formulae can be
derived from results in Heavy Meson ChPT \cite{Detmold:2006gh}.  In
Figure \ref{fig_chiral} we show the light quark mass
dependence for BSM operators.

\section{Summary}
We have given a status report on a computation of $\Delta S=2$ matrix elements
for a complete operator basis. The aim of this study is to compute the
renormalised matrix elements with dynamical fermions to have an impact on
phenomenological studies. 

The use of Domain Wall Fermions with very small chiral symmetry breaking is
essential for the continuum-like renormalisation of the operators. We use the
RI scheme with gauge-fixed momentum sources which greatly improves the
statistical accuracy of the renormalisation constants. We are exploring a
variant of the RI scheme with a non-exceptional momentum configuration to
reduce effects from chiral symmetry breaking at low momentum.

We have computed the bare matrix elements on the $32^3\times64\times16$
ensembles with small statistical errors for six valence masses. We plan to
extrapolate this data using SU(2) Chiral Perturbation Theory in the same way
as for $B_K$.

We also intend to extend this work to the coarser $24^3$ lattices
which will allow us to quantify the size of $O(a^2)$ lattice
artefacts.

\section*{Acknowledgements}
We thank our colleagues in RBC and UKQCD within whose programme this
calculation was performed. We thank the QCDOC design team for
developing the QCDOC machine and its software. This development and
the computers used in this calculation were funded by the U.S.DOE
grant DE-FG02-92ER40699, PPARC JIF grant PPA/J/S/1998/0075620 and by
RIKEN. This work was supported by DOE grant DE-FG02-92ER40699 and
PPARC grants PPA/G/O/2002/00465 and PP/D000238/1. We thank the
University of Edinburgh, PPARC, RIKEN, BNL and the U.S. DOE for
providing the QCDOC facilities used in this calculation.

\bibliography{fcnc}

\providecommand{\href}[2]{#2}\begingroup\raggedright\begin{thebibliography}{10}

\bibitem{Antonio:2007pb}
{\bf RBC and UKQCD} Collaboration, D.~J. Antonio {\em et~al.} {\em Phys. Rev.
  Lett.} {\bf 100} (2008) 032001
  [\href{http://arXiv.org/abs/hep-ph/0702042}{{\tt arXiv:hep-ph/0702042}}].

\bibitem{Aoki:2008ss}
{\bf JLQCD} Collaboration, S.~Aoki {\em et~al.} {\em Phys. Rev.} {\bf D77}
  (2008) 094503 [\href{http://arXiv.org/abs/0801.4186}{{\tt arXiv:0801.4186
  [hep-lat]}}].

\bibitem{Ciuchini:1998ix}
M.~Ciuchini {\em et~al.} {\em JHEP} {\bf 10} (1998) 008
  [\href{http://arXiv.org/abs/hep-ph/9808328}{{\tt arXiv:hep-ph/9808328}}].

\bibitem{Buras:2000if}
A.~J. Buras, M.~Misiak and J.~Urban {\em Nucl. Phys.} {\bf B586} (2000)
  397--426 [\href{http://arXiv.org/abs/hep-ph/0005183}{{\tt
  arXiv:hep-ph/0005183}}].

\bibitem{Martinelli:1994ty}
G.~Martinelli, C.~Pittori, C.~T. Sachrajda, M.~Testa and A.~Vladikas {\em Nucl.
  Phys.} {\bf B445} (1995) 81--108
  [\href{http://arXiv.org/abs/hep-lat/9411010}{{\tt arXiv:hep-lat/9411010}}].

\bibitem{Aoki:2007xm}
{\bf RBC and UKQCD} Collaboration, Y.~Aoki {\em et~al.}
  \href{http://arXiv.org/abs/0712.1061}{{\tt arXiv:0712.1061 [hep-lat]}}.

\bibitem{Babich:2006bh}
R.~Babich {\em et~al.} {\em Phys. Rev.} {\bf D74} (2006) 073009
  [\href{http://arXiv.org/abs/hep-lat/0605016}{{\tt arXiv:hep-lat/0605016}}].

\bibitem{Giusti:2000jr}
L.~Giusti and A.~Vladikas {\em Phys. Lett.} {\bf B488} (2000) 303--312
  [\href{http://arXiv.org/abs/hep-lat/0005026}{{\tt arXiv:hep-lat/0005026}}].

\bibitem{Becirevic:2004ny}
D.~Becirevic {\em et~al.} {\em JHEP} {\bf 08} (2004) 022
  [\href{http://arXiv.org/abs/hep-lat/0401033}{{\tt arXiv:hep-lat/0401033}}].

\bibitem{Aoki:2008xx}
{\bf RBC and UKQCD} Collaboration, Y.~Aoki {\em PoS} {\bf LAT2008} (2008) 222.

\bibitem{Scholz:2008uv}
{\bf RBC and UKQCD} Collaboration, E.~E. Scholz {\em PoS} {\bf LAT2008} (2008)
  095 [\href{http://arXiv.org/abs/0809.3251}{{\tt arXiv:0809.3251 [hep-lat]}}].

\bibitem{Kelly:2008aa}
{\bf RBC and UKQCD} Collaboration, C.~Kelly {\em PoS} {\bf LAT2008} (2008) 270.

\bibitem{Allton:2008pn}
{\bf RBC and UKQCD} Collaboration, C.~Allton {\em et~al.}
  \href{http://arXiv.org/abs/0804.0473}{{\tt arXiv:0804.0473 [hep-lat]}}.

\bibitem{Detmold:2006gh}
W.~Detmold and C.~J.~D. Lin {\em Phys. Rev.} {\bf D76} (2007) 014501
  [\href{http://arXiv.org/abs/hep-lat/0612028}{{\tt arXiv:hep-lat/0612028}}].

\end{thebibliography}\endgroup
\end{document}